\documentclass[aps,prl,twocolumn,showpacs]{revtex4}

\usepackage{subfigure}
\usepackage{graphicx}
\usepackage{bm}
\usepackage{amsmath}

\newcommand{\STO}{SrTiO$_3$}
\newcommand{\LAO}{LaAlO$_3$}

\begin{document}

\title{One-dimensional Quantum Wire Formed at the Boundary Between Two Insulating \LAO/\STO~ Interfaces}

\author{ A. Ron}
\affiliation{Raymond and Beverly Sackler School of Physics and Astronomy, Tel-Aviv University, Tel Aviv, 69978, Israel}
\author{Y. Dagan}
\affiliation{Raymond and Beverly Sackler School of Physics and Astronomy, Tel-Aviv University, Tel Aviv, 69978, Israel}

\begin{abstract}
We grow a tiled structure of insulating two dimensional LaAlO$_3$/SrTiO$_3$ interfaces composed of alternating one and three \LAO~ unit cells. The boundary between two tiles is conducting. At low temperatures this conductance exhibits quantized steps as a function of gate voltage indicative of a one dimensional channel. The step size of half the quantum of conductance is an evidence for absence of spin degeneracy.
\end{abstract}


\maketitle

Oxide interfaces can bring new functionalities into future electronic applications \cite{mannhart2010oxide}. Of particular interest are oxide based quantum wires that have tuanble spin-orbit interaction and superconductivity \cite{shalom2010tuning}, making them an important ingredient in future spin-based electronic devices, and in revealing exotic states, such as Majorana fermions \cite{oreg2010helical}. The hallmark interface between LaAlO$_3$ and SrTiO$_3$ surprisingly exhibits properties of a two-dimensional (2D) electron gas \cite{ohtomo2004high}, tunable superconductivity \cite{caviglia2008electric}, magnetism \cite{bert2011direct,li2011coexistence,dikin2011coexistence,brinkman2007magnetic,wang2011electronic} and tunable spin-orbit interaction \cite{shalom2010tuning,caviglia2010tunable}. Many attempts have been made to create a one dimensional (1D) channel out of the 2D LaAlO$_3/$SrTiO$_3$ interface, usually by narrowing down the 2D conductor using complex lithography techniques \cite{rakhmilevitch2010phase}. Here we present a new oxide 1D interface concept. This interface is formed between two 2D non-conducting interfaces of SrTiO$_3$ and LaAlO$_3$. We show that conducting electrons are confined to the boundary between these 2D interfaces in a potential well narrower than their Fermi wavelength. Consequently, they behave quantum-mechanically with their conductance appearing in discrete steps. From the step size we find that these electrons are spin polarized.
\par
\begin{figure}
 \includegraphics[width=1\hsize]{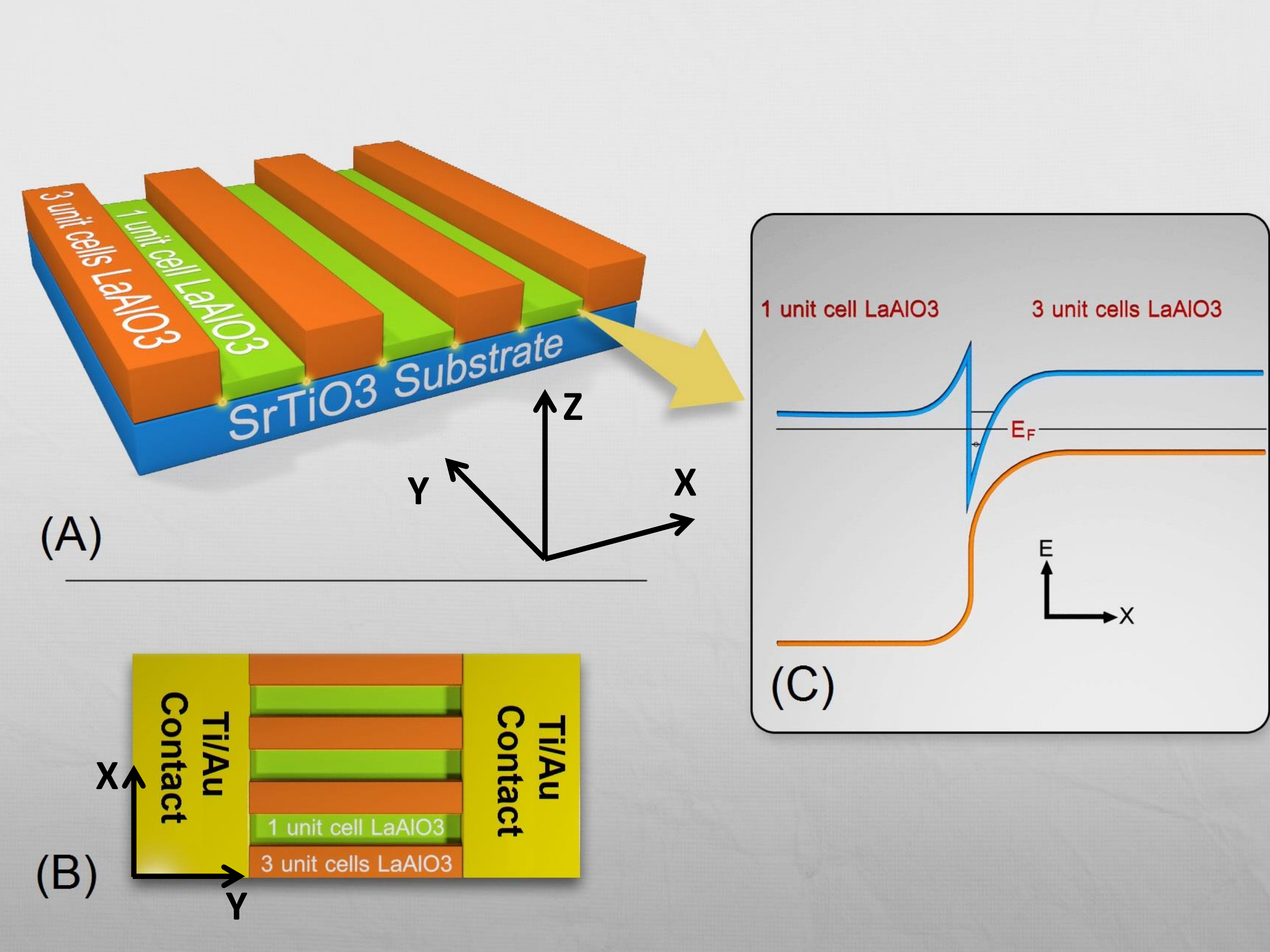}
  \caption{(color online). Figure 1.  Schematic description of the device. a Onto a TiO$_2$ terminated SrTiO$_3$ substrate we deposited sections of 1 unit cell LaAlO$_3$ adjacent to sections of 3 unit cell thick LaAlO$_3$, forming a 1D planar interface between them. The conducting wire is formed at this planar interface (highlighted in yellow). b Schematic illustration of the energy bands at the planar interface. Different thicknesses of the LaAlO$_3$ film have different energy gaps. At the planar interface, band bending occurs and a potential well is formed. This potential well hosts bound states, which become occupied when the Fermi energy is raised above them by applying gate voltage. c Schematic top view description of the device with Ti/Au ohmic contacts electrically connecting several 1D wires in parallel, the wires are placed 10$\mu$m apart.
}\label{Fig:1}
\end{figure}

Epitaxial films of \LAO~ are deposited using reflection high energy electron diffraction (RHEED) monitored pulsed laser deposition on atomically flat TiO$_2$ terminated \STO~ substrate in standard conditions \cite{shalom2009anisotropic}. In order to obtain the structure illustrated in figure \ref{Fig:1}a we first deposit a single unit cell thick layer of \LAO, then using standard photo-lithography technique, the planar interface is defined by depositing an amorphous oxide layer ,at room temperature, parallel to the crystal axis. \cite{schneider2006microlithography} followed by a liftoff step exposing the clean 1uc thick \LAO/\STO~ interface. Finally, a two unit cell thick layer of \LAO~  is deposited. This results in a planar interface between 1uc thick \LAO/\STO~ interface with an amorphous capping layer and a 3uc thick \LAO/\STO~ interface. A gold layer is evaporated as a back-gate. Ti-Au contacts are evaporated after Ar ion milling of the contact area. These contacts probe a few parallel boundaries as described in figure\ref{Fig:1}b. We use a wire bonder to connect voltage and current leads to each contact to eliminate the resistance of the leads, without eliminating the resistance between the metal and the planar interface. We ensured that the DC voltage drop across the sample was smaller than $k_B$T ($k_B$ being the Boltzmann constant and T is the temperature). The leakage current between the gate electrode and the device was always smaller than 1 nano-Ampere. Contact resistance is obtained by fitting the data to the Landauer formula with the contact resistance as a single fitting parameter


We verified that the layers themselves are insulating while the boundaries between them are conducting. No conductivity is expected neither at a 2D LaAlO$_3$/SrTiO$_3$ interface with a single nor with a triple unit cell thick LaAlO$_3$ layer used in our devices since both layers are below the threshold for conductivity for the 2D \STO/\LAO~ interfaces \cite{thiel2006tunable}. Furthermore, contacts that were put as close as 10$\mu$m away from the planar interface showed no conductance, suggesting that the conductivity is confined to a very narrow region. The resistance between these contacts is unmeasurably high and so is the resistance between two adjacent devices. The unexpected conductivity at the planar interface must stem from its discontinuity (see Figure 1 a, b). Our results show that a potential well is formed along the y direction (see Figure 1a).
\par
In Figure 2 the resistance as a function of temperature of one of these devices is shown. The overall behavior of the temperature dependence is similar to that observed in semiconductor heterojunctions \cite{pfeiffer1989electron}. The resistance increases upon cooling down, due to spurious carrier freeze-out. Further cooling results in a reduction of temperature-dependent scattering, such as electron-phonon. At low temperatures this process is overwhelmed by the depletion in the potential well formed. While the behavior described above is reproducible for each device, the details may vary from one sample to another. For example, the resistance extrema could appear at somewhat different temperatures for various devices. This may be related to random impurity states such as oxygen vacancies, La/Sr intermixing or other unknown impurity mechanisms.


\begin{figure}
 \includegraphics[width=1\hsize]{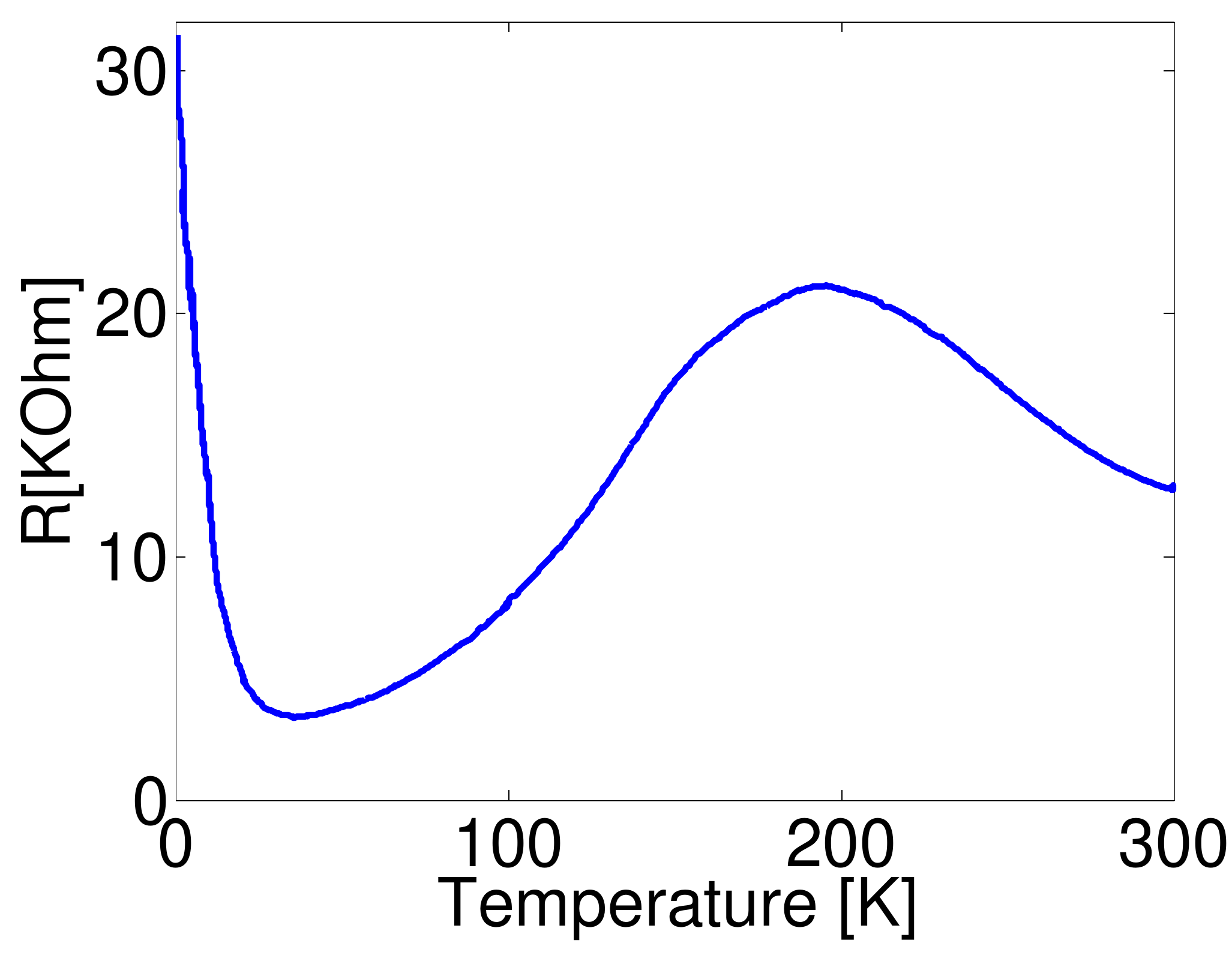}
  \caption{(color online).Figure 2.  Resistance versus temperature for 30 1 $\mu$m-long wires connected in parallel (at zero gate voltage using current of 1nA). The temperature dependence of the resistance exhibits the typical quantum confinement heterostructure behavior (see text for details).
}\label{Fig:2}
\end{figure}

\par
\begin{figure}
 \includegraphics[width=1\hsize]{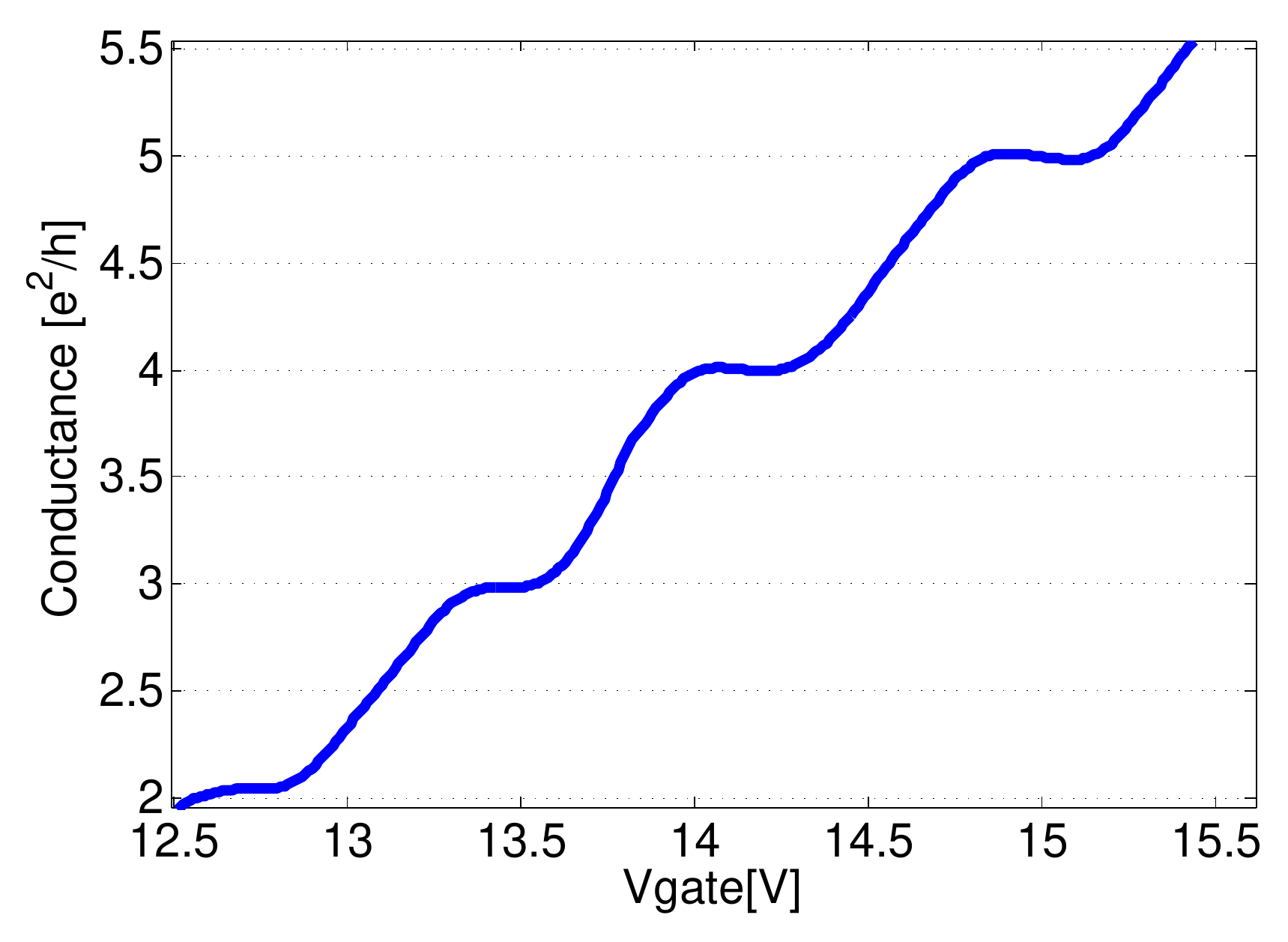}
  \caption{(color online).Figure 3. Conductance steps characteristics for 28 quantum wires 4$\mu$m-long. The wires are connected in parallel measured at 1.85 K after subtraction of serial contact resistance of 700$\Omega$. Clear conductance steps are seen with step height of $e^2/h$. Each step corresponds to a change of population in a single wire. For full voltage scan, see supplementary material\cite{Sup}.
}\label{Fig:3}
\end{figure}

Figure \ref{Fig:3} shows the conductance versus back-gate-voltage measured at 1.85 K for a $4\mu m $ long wire. Clear steps are observed. The step size $\frac{e^2}{h}$ is exactly half of the universal quantum conductance $g_0$ . This is indicative of the existence of a 1D confinement with quantized energy levels.
\par


According to R. Landauer \cite{Imry} the conductance in a 1D channel is quantized and can be expressed as
$$ g=\sum_{i}T_ig_0$$
Here $T_i$ are the transmission coefficients of the various channels, and $g_0=\frac{2e^2}{h}$ is the quantum conductance with $e$ the elementary charge and $h$ the Planck constant. For a ballistic device, whose length is shorter than the mean free path $l$ the transmissions, $T_i=1$ and $$\sum_{i}T_i=N$$, where $N$ is an integer determined by the number of occupied modes.
\par
The devices are comprised of a few parallel wires (figure \ref{Fig:1}b). Such parallel measurement can only increase the conductance relative to a single wire. If the population of two wires is increased within the same gate voltage range, the conductance will increase by a step twice as high as the usual one. (See supplementary material\cite{Sup} for demonstration of such an occurrence).
\par

Conductance steps lower than $g_0$ can be observed when the transmission coefficients $T_i$ are smaller than unity, usually due to scattering along the wire \cite{Imry}. This is not the case in our device. To eliminate the low transmission coefficient scenario, we fabricate a shorter wire ($1\mu m$ length) and a longer one ($30\mu m$). It is expected that the scattering probability is increased with wire length. However, we observe a reproducible $\frac{g_0}{2}$ step height for both $1\mu m$ (see Figure 4a) and $4\mu m$ wires. A diffusive channel ($T_i < 1$) is obtained only for the longer wire (see Figure 4b). This indicates a completely ballistic process for the shorter wires of up to $4\mu m$. Furthermore, in the diffusive device, the step height is no longer uniform, as expected.
\par
Another possible mechanism for $\frac{g_0}{2}$ conductance is the removal of spin degeneracy. This has been demonstrated to happen under applied magnetic field \cite{quay2010observation}. In the absence of an external magnetic field, this is possible if the conducting electrons are spin-polarized \cite{ono1999equation, Wigner}.

Magnetic effects have been observed in the 2D interface between LaAlO$_3$ and SrTiO$_3$ using various probes \cite{bert2011direct,li2011coexistence,dikin2011coexistence,brinkman2007magnetic,lee2013titanium,kalisky2012critical}. This magnetism has been related to the polar nature of this interface \cite{flekser2012magnetotransport} appearing only above a critical thickness of four unit cells \cite{kalisky2012critical}. It is therefore difficult to confirm whether the magnetic effects in 2D and the spin polarization reported here are of the same origin.
\par

\begin{figure}
 \includegraphics[width=1\hsize]{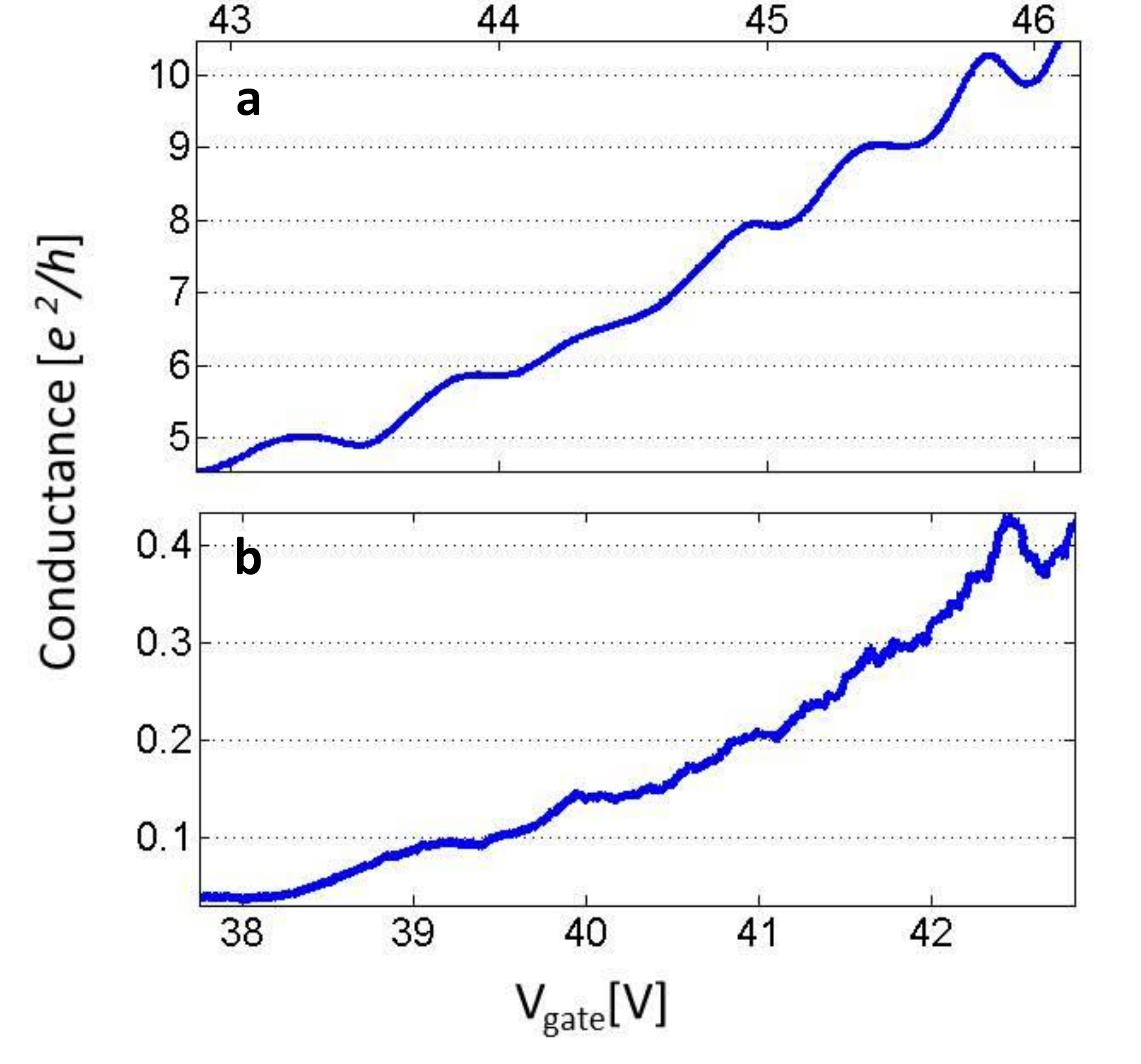}
  \caption{(color online).Figure 4 Conductance versus back gate voltage for various wire lengths a. 30 wires 1$\mu$ m-long connected in parallel at 0.55 K after subtraction of serial contact resistance of 3.5 K.  b. 20 wires 30$\mu$m-long connected in parallel at 2 K. These wires are longer than the ballistic length, which results in smaller and non-uniform step heights.
}\label{Fig:4}
\end{figure}

One may claim that the quantized conductance steps arise from a point contact connected in series to a 2D reservoir \cite{van1988quantized}. We rule out this possibility in our devices. A 2D conductor made of LaAlO$_3$/SrTiO$_3$ connected in series to a point contact would have a highly gate voltage-sensitive resistance. For example, for the gate voltage range studied here (see Figure 3 and supplementary material\cite{Sup}), one would expect the resistance of the presumed 2D section to change by a factor of 2.6 \cite{shalom2010tuning}. This would result in a noticeable deviation from the universal step values observed. We therefore conclude that our device is a quantum wire.
\par
We now discuss possible mechanisms for the formation of the quantum wire at the 1D interface.
While the origin of the 2D conductivity in the LaAlO$_3$/SrTiO$_3$ interface is still a matter of strong debate it has been established that conductivity appears only above a threshold of four \LAO~ layers \cite{thiel2006tunable}. A simplified electrostatic consideration suggests that adding LaAlO$_3$ layers results in the accumulation of electrostatic potential energy. When this energy becomes of the order of the band gap of SrTiO$_3$ conducting electrons are transferred into the interface.
\par
A more detailed calculation showed that below the conductivity threshold \cite{thiel2006tunable} the gap between the unoccupied (titanium 3d states) and occupied bands (oxygen 2p states)~monotonically decreases upon accumulation of LaAlO$_3$ layers \cite{pentcheva2006charge}.  Adopting this view the gap between the unoccupied and the occupied bands changes discontinuously across the 1D interface when the \LAO~ layer thickness suddenly changes (say, along the x direction see fig\ref{Fig:1}a). Consequently, a potential well should be formed (Figure 1c). Such discontinuity is similar to that implemented in semiconductor heterostructures with different band gaps (e.g. GaAs/Al$_x$Ga$_{1-x}$As). There, the band discontinuity is along the z axis resulting in a 2D potential well formed between the two semiconductors. Although in the semiconductor heterostructure charge modulation doping layer away from the interface is usually implanted in our case the origin of the initial charges is not clear to us. The fact that our interface is placed only 12$\AA$ away from the surface and on top of \STO~ substrate, which has many doping mechanisms may explain the existence of charges as seen in the temperature dependence of our device.
\par
It has been recently reported that the surface potential in LaAlO$_3$/SrTiO$_3$ interface has a stripe-like structure. These stripes are related to the domain structure of SrTiO$_3$ below the cubic to tetragonal transition. \cite{kalisky2013locally} It is possible that a domain wall is pinned to our planar interface resulting in a higher charge concentration and better mobility. However, this is not enough to explain the conductivity in the wire since for 3 unit cell thick LaAlO$_3$ layer on top of SrTiO$_3$ no conductivity is observed despite the domains. A detailed calculation taking into account the polar layers and the presence of domain wall is needed to explore this possibility.
\par
The analysis of the conductance versus gate voltage in figures 3,4a yields a transmission coefficients very close to unity. i.e. the wires are ballistic. This is in spite of their lengths of 4$\mu m$ and 1$\mu m$ respectively. This is much longer than the mean free path of the order of $0.1\mu m$, measured for the LaAlO$_3$/SrTiO$_3$ 2D interface \cite{shalom2009anisotropic}.
\par
Anomalous high mobility has recently been reported in SrTiO$_3$/LaAlO$_3$ nano-wires, defined using an atomic force microscope with a conducting tip \cite{irvin2013anomalous}. A mean free path enhancement in a 1D system, compared to a 2D case, is predicted due to a reduced phase space available for back-scattering \cite{sakaki1980scattering}. However, in semiconductors this enhancement is usually not large \cite{kaufman1999conductance}. We conjecture that the long-ranged ballistic length in our case is a result of the polarized electrons experiencing spin-orbit interaction \cite{shalom2010tuning,caviglia2010tunable}. In such a scenario, backscattering is strongly suppressed, since reversing momentum direction requires spin-flip scattering and surmounting the exchange energy.
\par
This letter presents a novel and simple method of fabricating a one-dimensional quantum wire, utilizing the polar structure of the LaAlO$_3$/SrTiO$_3$ interface. This new wire opens possibilities for long-ranged spin transport. Furthermore, the constituting oxides will introduce new functionalities and correlations into this wire \cite{oreg2013fractional}. The possibility to control correlation effects, spin-orbit interactions and coupling to superconductivity makes this wire a laboratory for studying Majorana fermions \cite{oreg2010helical} and fractional charges \cite{oreg2013fractional}.

We thank Eran Sela, Alexander Palevski and Moshe Ben Shalom for useful discussions.
This work was supported in part by the Israeli Science Foundation under grant no.569/13 by the Ministry of Science and Technology under contract 3-8667 and by the US-Israel bi-national science foundation (BSF).

\bibliographystyle{apsrev}
\bibliography{myBib2}

\begin{thebibliography}{32}
\expandafter\ifx\csname natexlab\endcsname\relax\def\natexlab#1{#1}\fi
\expandafter\ifx\csname bibnamefont\endcsname\relax
  \def\bibnamefont#1{#1}\fi
\expandafter\ifx\csname bibfnamefont\endcsname\relax
  \def\bibfnamefont#1{#1}\fi
\expandafter\ifx\csname citenamefont\endcsname\relax
  \def\citenamefont#1{#1}\fi
\expandafter\ifx\csname url\endcsname\relax
  \def\url#1{\texttt{#1}}\fi
\expandafter\ifx\csname urlprefix\endcsname\relax\def\urlprefix{URL }\fi
\providecommand{\bibinfo}[2]{#2}
\providecommand{\eprint}[2][]{\url{#2}}

\bibitem[{\citenamefont{Mannhart and Schlom}(2010)}]{mannhart2010oxide}
\bibinfo{author}{\bibfnamefont{J.}~\bibnamefont{Mannhart}} \bibnamefont{and}
  \bibinfo{author}{\bibfnamefont{D.}~\bibnamefont{Schlom}},
  \bibinfo{journal}{Science} \textbf{\bibinfo{volume}{327}},
  \bibinfo{pages}{1607} (\bibinfo{year}{2010}).

\bibitem[{\citenamefont{Ben~Shalom et~al.}(2010)\citenamefont{Ben~Shalom,
  Sachs, Rakhmilevitch, Palevski, and Dagan}}]{shalom2010tuning}
\bibinfo{author}{\bibfnamefont{M.}~\bibnamefont{Ben~Shalom}},
  \bibinfo{author}{\bibfnamefont{M.}~\bibnamefont{Sachs}},
  \bibinfo{author}{\bibfnamefont{D.}~\bibnamefont{Rakhmilevitch}},
  \bibinfo{author}{\bibfnamefont{A.}~\bibnamefont{Palevski}}, \bibnamefont{and}
  \bibinfo{author}{\bibfnamefont{Y.}~\bibnamefont{Dagan}},
  \bibinfo{journal}{Physical review letters} \textbf{\bibinfo{volume}{104}},
  \bibinfo{pages}{126802} (\bibinfo{year}{2010}).

\bibitem[{\citenamefont{Oreg et~al.}(2010)\citenamefont{Oreg, Refael, and von
  Oppen}}]{oreg2010helical}
\bibinfo{author}{\bibfnamefont{Y.}~\bibnamefont{Oreg}},
  \bibinfo{author}{\bibfnamefont{G.}~\bibnamefont{Refael}}, \bibnamefont{and}
  \bibinfo{author}{\bibfnamefont{F.}~\bibnamefont{von Oppen}},
  \bibinfo{journal}{Physical review letters} \textbf{\bibinfo{volume}{105}},
  \bibinfo{pages}{177002} (\bibinfo{year}{2010}).

\bibitem[{\citenamefont{Ohtomo and Hwang}(2004)}]{ohtomo2004high}
\bibinfo{author}{\bibfnamefont{A.}~\bibnamefont{Ohtomo}} \bibnamefont{and}
  \bibinfo{author}{\bibfnamefont{H.}~\bibnamefont{Hwang}},
  \bibinfo{journal}{Nature} \textbf{\bibinfo{volume}{427}},
  \bibinfo{pages}{423} (\bibinfo{year}{2004}).

\bibitem[{\citenamefont{Caviglia et~al.}(2008)\citenamefont{Caviglia, Gariglio,
  Reyren, Jaccard, Schneider, Gabay, Thiel, Hammerl, Mannhart, and
  Triscone}}]{caviglia2008electric}
\bibinfo{author}{\bibfnamefont{A.}~\bibnamefont{Caviglia}},
  \bibinfo{author}{\bibfnamefont{S.}~\bibnamefont{Gariglio}},
  \bibinfo{author}{\bibfnamefont{N.}~\bibnamefont{Reyren}},
  \bibinfo{author}{\bibfnamefont{D.}~\bibnamefont{Jaccard}},
  \bibinfo{author}{\bibfnamefont{T.}~\bibnamefont{Schneider}},
  \bibinfo{author}{\bibfnamefont{M.}~\bibnamefont{Gabay}},
  \bibinfo{author}{\bibfnamefont{S.}~\bibnamefont{Thiel}},
  \bibinfo{author}{\bibfnamefont{G.}~\bibnamefont{Hammerl}},
  \bibinfo{author}{\bibfnamefont{J.}~\bibnamefont{Mannhart}}, \bibnamefont{and}
  \bibinfo{author}{\bibfnamefont{J.-M.} \bibnamefont{Triscone}},
  \bibinfo{journal}{Nature} \textbf{\bibinfo{volume}{456}},
  \bibinfo{pages}{624} (\bibinfo{year}{2008}).

\bibitem[{\citenamefont{Bert et~al.}(2011)\citenamefont{Bert, Kalisky, Bell,
  Kim, Hikita, Hwang, and Moler}}]{bert2011direct}
\bibinfo{author}{\bibfnamefont{J.~A.} \bibnamefont{Bert}},
  \bibinfo{author}{\bibfnamefont{B.}~\bibnamefont{Kalisky}},
  \bibinfo{author}{\bibfnamefont{C.}~\bibnamefont{Bell}},
  \bibinfo{author}{\bibfnamefont{M.}~\bibnamefont{Kim}},
  \bibinfo{author}{\bibfnamefont{Y.}~\bibnamefont{Hikita}},
  \bibinfo{author}{\bibfnamefont{H.~Y.} \bibnamefont{Hwang}}, \bibnamefont{and}
  \bibinfo{author}{\bibfnamefont{K.~A.} \bibnamefont{Moler}},
  \bibinfo{journal}{Nature physics} \textbf{\bibinfo{volume}{7}},
  \bibinfo{pages}{767} (\bibinfo{year}{2011}).

\bibitem[{\citenamefont{Li et~al.}(2011)\citenamefont{Li, Richter, Mannhart,
  and Ashoori}}]{li2011coexistence}
\bibinfo{author}{\bibfnamefont{L.}~\bibnamefont{Li}},
  \bibinfo{author}{\bibfnamefont{C.}~\bibnamefont{Richter}},
  \bibinfo{author}{\bibfnamefont{J.}~\bibnamefont{Mannhart}}, \bibnamefont{and}
  \bibinfo{author}{\bibfnamefont{R.}~\bibnamefont{Ashoori}},
  \bibinfo{journal}{Nature Physics} \textbf{\bibinfo{volume}{7}},
  \bibinfo{pages}{762} (\bibinfo{year}{2011}).

\bibitem[{\citenamefont{Dikin et~al.}(2011)\citenamefont{Dikin, Mehta, Bark,
  Folkman, Eom, and Chandrasekhar}}]{dikin2011coexistence}
\bibinfo{author}{\bibfnamefont{D.}~\bibnamefont{Dikin}},
  \bibinfo{author}{\bibfnamefont{M.}~\bibnamefont{Mehta}},
  \bibinfo{author}{\bibfnamefont{C.}~\bibnamefont{Bark}},
  \bibinfo{author}{\bibfnamefont{C.}~\bibnamefont{Folkman}},
  \bibinfo{author}{\bibfnamefont{C.}~\bibnamefont{Eom}}, \bibnamefont{and}
  \bibinfo{author}{\bibfnamefont{V.}~\bibnamefont{Chandrasekhar}},
  \bibinfo{journal}{Physical Review Letters} \textbf{\bibinfo{volume}{107}},
  \bibinfo{pages}{056802} (\bibinfo{year}{2011}).

\bibitem[{\citenamefont{Brinkman et~al.}(2007)\citenamefont{Brinkman, Huijben,
  Van~Zalk, Huijben, Zeitler, Maan, Van~der Wiel, Rijnders, Blank, and
  Hilgenkamp}}]{brinkman2007magnetic}
\bibinfo{author}{\bibfnamefont{A.}~\bibnamefont{Brinkman}},
  \bibinfo{author}{\bibfnamefont{M.}~\bibnamefont{Huijben}},
  \bibinfo{author}{\bibfnamefont{M.}~\bibnamefont{Van~Zalk}},
  \bibinfo{author}{\bibfnamefont{J.}~\bibnamefont{Huijben}},
  \bibinfo{author}{\bibfnamefont{U.}~\bibnamefont{Zeitler}},
  \bibinfo{author}{\bibfnamefont{J.}~\bibnamefont{Maan}},
  \bibinfo{author}{\bibfnamefont{W.}~\bibnamefont{Van~der Wiel}},
  \bibinfo{author}{\bibfnamefont{G.}~\bibnamefont{Rijnders}},
  \bibinfo{author}{\bibfnamefont{D.}~\bibnamefont{Blank}}, \bibnamefont{and}
  \bibinfo{author}{\bibfnamefont{H.}~\bibnamefont{Hilgenkamp}},
  \bibinfo{journal}{Nature materials} \textbf{\bibinfo{volume}{6}},
  \bibinfo{pages}{493} (\bibinfo{year}{2007}).

\bibitem[{\citenamefont{Wang et~al.}(2011)\citenamefont{Wang, Baskaran, Liu,
  Huijben, Yi, Annadi, Barman, Rusydi, Dhar, Feng et~al.}}]{wang2011electronic}
\bibinfo{author}{\bibfnamefont{X.}~\bibnamefont{Wang}},
  \bibinfo{author}{\bibfnamefont{G.}~\bibnamefont{Baskaran}},
  \bibinfo{author}{\bibfnamefont{Z.}~\bibnamefont{Liu}},
  \bibinfo{author}{\bibfnamefont{J.}~\bibnamefont{Huijben}},
  \bibinfo{author}{\bibfnamefont{J.}~\bibnamefont{Yi}},
  \bibinfo{author}{\bibfnamefont{A.}~\bibnamefont{Annadi}},
  \bibinfo{author}{\bibfnamefont{A.~R.} \bibnamefont{Barman}},
  \bibinfo{author}{\bibfnamefont{A.}~\bibnamefont{Rusydi}},
  \bibinfo{author}{\bibfnamefont{S.}~\bibnamefont{Dhar}},
  \bibinfo{author}{\bibfnamefont{Y.}~\bibnamefont{Feng}}, \bibnamefont{et~al.},
  \bibinfo{journal}{Nature Communications} \textbf{\bibinfo{volume}{2}},
  \bibinfo{pages}{188} (\bibinfo{year}{2011}).

\bibitem[{\citenamefont{Caviglia et~al.}(2010)\citenamefont{Caviglia, Gabay,
  Gariglio, Reyren, Cancellieri, and Triscone}}]{caviglia2010tunable}
\bibinfo{author}{\bibfnamefont{A.}~\bibnamefont{Caviglia}},
  \bibinfo{author}{\bibfnamefont{M.}~\bibnamefont{Gabay}},
  \bibinfo{author}{\bibfnamefont{S.}~\bibnamefont{Gariglio}},
  \bibinfo{author}{\bibfnamefont{N.}~\bibnamefont{Reyren}},
  \bibinfo{author}{\bibfnamefont{C.}~\bibnamefont{Cancellieri}},
  \bibnamefont{and} \bibinfo{author}{\bibfnamefont{J.-M.}
  \bibnamefont{Triscone}}, \bibinfo{journal}{Physical review letters}
  \textbf{\bibinfo{volume}{104}}, \bibinfo{pages}{126803}
  (\bibinfo{year}{2010}).

\bibitem[{\citenamefont{Rakhmilevitch et~al.}(2010)\citenamefont{Rakhmilevitch,
  Ben~Shalom, Eshkol, Tsukernik, Palevski, and Dagan}}]{rakhmilevitch2010phase}
\bibinfo{author}{\bibfnamefont{D.}~\bibnamefont{Rakhmilevitch}},
  \bibinfo{author}{\bibfnamefont{M.}~\bibnamefont{Ben~Shalom}},
  \bibinfo{author}{\bibfnamefont{M.}~\bibnamefont{Eshkol}},
  \bibinfo{author}{\bibfnamefont{A.}~\bibnamefont{Tsukernik}},
  \bibinfo{author}{\bibfnamefont{A.}~\bibnamefont{Palevski}}, \bibnamefont{and}
  \bibinfo{author}{\bibfnamefont{Y.}~\bibnamefont{Dagan}},
  \bibinfo{journal}{Physical Review B} \textbf{\bibinfo{volume}{82}},
  \bibinfo{pages}{235119} (\bibinfo{year}{2010}).

\bibitem[{\citenamefont{Ben~Shalom et~al.}(2009)\citenamefont{Ben~Shalom, Tai,
  Lereah, Sachs, Levy, Rakhmilevitch, Palevski, and
  Dagan}}]{shalom2009anisotropic}
\bibinfo{author}{\bibfnamefont{M.}~\bibnamefont{Ben~Shalom}},
  \bibinfo{author}{\bibfnamefont{C.}~\bibnamefont{Tai}},
  \bibinfo{author}{\bibfnamefont{Y.}~\bibnamefont{Lereah}},
  \bibinfo{author}{\bibfnamefont{M.}~\bibnamefont{Sachs}},
  \bibinfo{author}{\bibfnamefont{E.}~\bibnamefont{Levy}},
  \bibinfo{author}{\bibfnamefont{D.}~\bibnamefont{Rakhmilevitch}},
  \bibinfo{author}{\bibfnamefont{A.}~\bibnamefont{Palevski}}, \bibnamefont{and}
  \bibinfo{author}{\bibfnamefont{Y.}~\bibnamefont{Dagan}},
  \bibinfo{journal}{Physical Review B} \textbf{\bibinfo{volume}{80}},
  \bibinfo{pages}{140403} (\bibinfo{year}{2009}).

\bibitem[{\citenamefont{Schneider et~al.}(2006)\citenamefont{Schneider, Thiel,
  Hammerl, Richter, and Mannhart}}]{schneider2006microlithography}
\bibinfo{author}{\bibfnamefont{C.}~\bibnamefont{Schneider}},
  \bibinfo{author}{\bibfnamefont{S.}~\bibnamefont{Thiel}},
  \bibinfo{author}{\bibfnamefont{G.}~\bibnamefont{Hammerl}},
  \bibinfo{author}{\bibfnamefont{C.}~\bibnamefont{Richter}}, \bibnamefont{and}
  \bibinfo{author}{\bibfnamefont{J.}~\bibnamefont{Mannhart}},
  \bibinfo{journal}{Applied physics letters} \textbf{\bibinfo{volume}{89}},
  \bibinfo{pages}{122101} (\bibinfo{year}{2006}).

\bibitem[{\citenamefont{Thiel et~al.}(2006)\citenamefont{Thiel, Hammerl,
  Schmehl, Schneider, and Mannhart}}]{thiel2006tunable}
\bibinfo{author}{\bibfnamefont{S.}~\bibnamefont{Thiel}},
  \bibinfo{author}{\bibfnamefont{G.}~\bibnamefont{Hammerl}},
  \bibinfo{author}{\bibfnamefont{A.}~\bibnamefont{Schmehl}},
  \bibinfo{author}{\bibfnamefont{C.}~\bibnamefont{Schneider}},
  \bibnamefont{and} \bibinfo{author}{\bibfnamefont{J.}~\bibnamefont{Mannhart}},
  \bibinfo{journal}{Science} \textbf{\bibinfo{volume}{313}},
  \bibinfo{pages}{1942} (\bibinfo{year}{2006}).

\bibitem[{\citenamefont{Pfeiffer et~al.}(1989)\citenamefont{Pfeiffer, West,
  Stormer, and Baldwin}}]{pfeiffer1989electron}
\bibinfo{author}{\bibfnamefont{L.}~\bibnamefont{Pfeiffer}},
  \bibinfo{author}{\bibfnamefont{K.}~\bibnamefont{West}},
  \bibinfo{author}{\bibfnamefont{H.}~\bibnamefont{Stormer}}, \bibnamefont{and}
  \bibinfo{author}{\bibfnamefont{K.}~\bibnamefont{Baldwin}},
  \bibinfo{journal}{Applied Physics Letters} \textbf{\bibinfo{volume}{55}},
  \bibinfo{pages}{1888} (\bibinfo{year}{1989}).

\bibitem[{Sup()}]{Sup}
\bibinfo{note}{See Supplemental Material online, which includes Ref.
  \cite{biscaras2012irreversibility}}.

\bibitem[{\citenamefont{Imry}(2002)}]{Imry}
\bibinfo{author}{\bibfnamefont{Y.}~\bibnamefont{Imry}},
  \emph{\bibinfo{title}{Introduction to mesoscopic physics (Vol. 2)}}
  (\bibinfo{publisher}{Oxford University Press on Demand},
  \bibinfo{address}{Oxford}, \bibinfo{year}{2002}).

\bibitem[{\citenamefont{Quay et~al.}(2010)\citenamefont{Quay, Hughes, Sulpizio,
  Pfeiffer, Baldwin, West, Goldhaber-Gordon, and
  De~Picciotto}}]{quay2010observation}
\bibinfo{author}{\bibfnamefont{C.}~\bibnamefont{Quay}},
  \bibinfo{author}{\bibfnamefont{T.}~\bibnamefont{Hughes}},
  \bibinfo{author}{\bibfnamefont{J.}~\bibnamefont{Sulpizio}},
  \bibinfo{author}{\bibfnamefont{L.}~\bibnamefont{Pfeiffer}},
  \bibinfo{author}{\bibfnamefont{K.}~\bibnamefont{Baldwin}},
  \bibinfo{author}{\bibfnamefont{K.}~\bibnamefont{West}},
  \bibinfo{author}{\bibfnamefont{D.}~\bibnamefont{Goldhaber-Gordon}},
  \bibnamefont{and}
  \bibinfo{author}{\bibfnamefont{R.}~\bibnamefont{De~Picciotto}},
  \bibinfo{journal}{Nature Physics} \textbf{\bibinfo{volume}{6}},
  \bibinfo{pages}{336} (\bibinfo{year}{2010}).

\bibitem[{\citenamefont{Ono et~al.}(1999)\citenamefont{Ono, Ooka, Miyajima, and
  Otani}}]{ono1999equation}
\bibinfo{author}{\bibfnamefont{T.}~\bibnamefont{Ono}},
  \bibinfo{author}{\bibfnamefont{Y.}~\bibnamefont{Ooka}},
  \bibinfo{author}{\bibfnamefont{H.}~\bibnamefont{Miyajima}}, \bibnamefont{and}
  \bibinfo{author}{\bibfnamefont{Y.}~\bibnamefont{Otani}},
  \bibinfo{journal}{Applied Physics Letters} \textbf{\bibinfo{volume}{75}},
  \bibinfo{pages}{1622} (\bibinfo{year}{1999}).

\bibitem[{Wig()}]{Wigner}
\bibinfo{note}{Wigner crystal can also result in e2/h conductance step height.
  However, due to the large dielectric constant in SrTiO$_3$ the expected
  density for a Wigner crystal is about 4 electrons per micron length. This is
  unreasonably small for the gate voltage studied.}

\bibitem[{\citenamefont{Lee et~al.}(2013)\citenamefont{Lee, Xie, Sato, Bell,
  Hikita, Hwang, and Kao}}]{lee2013titanium}
\bibinfo{author}{\bibfnamefont{J.-S.} \bibnamefont{Lee}},
  \bibinfo{author}{\bibfnamefont{Y.}~\bibnamefont{Xie}},
  \bibinfo{author}{\bibfnamefont{H.}~\bibnamefont{Sato}},
  \bibinfo{author}{\bibfnamefont{C.}~\bibnamefont{Bell}},
  \bibinfo{author}{\bibfnamefont{Y.}~\bibnamefont{Hikita}},
  \bibinfo{author}{\bibfnamefont{H.}~\bibnamefont{Hwang}}, \bibnamefont{and}
  \bibinfo{author}{\bibfnamefont{C.-C.} \bibnamefont{Kao}},
  \bibinfo{journal}{Nature materials}  (\bibinfo{year}{2013}).

\bibitem[{\citenamefont{Kalisky et~al.}(2012)\citenamefont{Kalisky, Bert,
  Klopfer, Bell, Sato, Hosoda, Hikita, Hwang, and Moler}}]{kalisky2012critical}
\bibinfo{author}{\bibfnamefont{B.}~\bibnamefont{Kalisky}},
  \bibinfo{author}{\bibfnamefont{J.~A.} \bibnamefont{Bert}},
  \bibinfo{author}{\bibfnamefont{B.~B.} \bibnamefont{Klopfer}},
  \bibinfo{author}{\bibfnamefont{C.}~\bibnamefont{Bell}},
  \bibinfo{author}{\bibfnamefont{H.~K.} \bibnamefont{Sato}},
  \bibinfo{author}{\bibfnamefont{M.}~\bibnamefont{Hosoda}},
  \bibinfo{author}{\bibfnamefont{Y.}~\bibnamefont{Hikita}},
  \bibinfo{author}{\bibfnamefont{H.~Y.} \bibnamefont{Hwang}}, \bibnamefont{and}
  \bibinfo{author}{\bibfnamefont{K.~A.} \bibnamefont{Moler}},
  \bibinfo{journal}{Nature Communications} \textbf{\bibinfo{volume}{3}},
  \bibinfo{pages}{922} (\bibinfo{year}{2012}).

\bibitem[{\citenamefont{Flekser et~al.}(2012)\citenamefont{Flekser, Ben~Shalom,
  Kim, Bell, Hikita, Hwang, and Dagan}}]{flekser2012magnetotransport}
\bibinfo{author}{\bibfnamefont{E.}~\bibnamefont{Flekser}},
  \bibinfo{author}{\bibfnamefont{M.}~\bibnamefont{Ben~Shalom}},
  \bibinfo{author}{\bibfnamefont{M.}~\bibnamefont{Kim}},
  \bibinfo{author}{\bibfnamefont{C.}~\bibnamefont{Bell}},
  \bibinfo{author}{\bibfnamefont{Y.}~\bibnamefont{Hikita}},
  \bibinfo{author}{\bibfnamefont{H.}~\bibnamefont{Hwang}}, \bibnamefont{and}
  \bibinfo{author}{\bibfnamefont{Y.}~\bibnamefont{Dagan}},
  \bibinfo{journal}{Physical Review B} \textbf{\bibinfo{volume}{86}},
  \bibinfo{pages}{121104} (\bibinfo{year}{2012}).

\bibitem[{\citenamefont{Van~Wees et~al.}(1988)\citenamefont{Van~Wees,
  Van~Houten, Beenakker, Williamson, Kouwenhoven, Van~der Marel, and
  Foxon}}]{van1988quantized}
\bibinfo{author}{\bibfnamefont{B.}~\bibnamefont{Van~Wees}},
  \bibinfo{author}{\bibfnamefont{H.}~\bibnamefont{Van~Houten}},
  \bibinfo{author}{\bibfnamefont{C.}~\bibnamefont{Beenakker}},
  \bibinfo{author}{\bibfnamefont{J.~G.} \bibnamefont{Williamson}},
  \bibinfo{author}{\bibfnamefont{L.}~\bibnamefont{Kouwenhoven}},
  \bibinfo{author}{\bibfnamefont{D.}~\bibnamefont{Van~der Marel}},
  \bibnamefont{and} \bibinfo{author}{\bibfnamefont{C.}~\bibnamefont{Foxon}},
  \bibinfo{journal}{Physical Review Letters} \textbf{\bibinfo{volume}{60}},
  \bibinfo{pages}{848} (\bibinfo{year}{1988}).

\bibitem[{\citenamefont{Pentcheva and Pickett}(2006)}]{pentcheva2006charge}
\bibinfo{author}{\bibfnamefont{R.}~\bibnamefont{Pentcheva}} \bibnamefont{and}
  \bibinfo{author}{\bibfnamefont{W.}~\bibnamefont{Pickett}},
  \bibinfo{journal}{Physical Review B} \textbf{\bibinfo{volume}{74}},
  \bibinfo{pages}{035112} (\bibinfo{year}{2006}).

\bibitem[{\citenamefont{Kalisky et~al.}(2013)\citenamefont{Kalisky, Spanton,
  Noad, Kirtley, Nowack, Bell, Sato, Hosoda, Xie, Hikita
  et~al.}}]{kalisky2013locally}
\bibinfo{author}{\bibfnamefont{B.}~\bibnamefont{Kalisky}},
  \bibinfo{author}{\bibfnamefont{E.~M.} \bibnamefont{Spanton}},
  \bibinfo{author}{\bibfnamefont{H.}~\bibnamefont{Noad}},
  \bibinfo{author}{\bibfnamefont{J.~R.} \bibnamefont{Kirtley}},
  \bibinfo{author}{\bibfnamefont{K.~C.} \bibnamefont{Nowack}},
  \bibinfo{author}{\bibfnamefont{C.}~\bibnamefont{Bell}},
  \bibinfo{author}{\bibfnamefont{H.~K.} \bibnamefont{Sato}},
  \bibinfo{author}{\bibfnamefont{M.}~\bibnamefont{Hosoda}},
  \bibinfo{author}{\bibfnamefont{Y.}~\bibnamefont{Xie}},
  \bibinfo{author}{\bibfnamefont{Y.}~\bibnamefont{Hikita}},
  \bibnamefont{et~al.}, \bibinfo{journal}{Nature materials}
  (\bibinfo{year}{2013}).

\bibitem[{\citenamefont{Irvin et~al.}(2013)\citenamefont{Irvin, Veazey, Cheng,
  Lu, Bark, Ryu, Eom, and Levy}}]{irvin2013anomalous}
\bibinfo{author}{\bibfnamefont{P.}~\bibnamefont{Irvin}},
  \bibinfo{author}{\bibfnamefont{J.~P.} \bibnamefont{Veazey}},
  \bibinfo{author}{\bibfnamefont{G.}~\bibnamefont{Cheng}},
  \bibinfo{author}{\bibfnamefont{S.}~\bibnamefont{Lu}},
  \bibinfo{author}{\bibfnamefont{C.-W.} \bibnamefont{Bark}},
  \bibinfo{author}{\bibfnamefont{S.}~\bibnamefont{Ryu}},
  \bibinfo{author}{\bibfnamefont{C.-B.} \bibnamefont{Eom}}, \bibnamefont{and}
  \bibinfo{author}{\bibfnamefont{J.}~\bibnamefont{Levy}},
  \bibinfo{journal}{Nano letters} \textbf{\bibinfo{volume}{13}},
  \bibinfo{pages}{364} (\bibinfo{year}{2013}).

\bibitem[{\citenamefont{Sakaki}(1980)}]{sakaki1980scattering}
\bibinfo{author}{\bibfnamefont{H.}~\bibnamefont{Sakaki}},
  \bibinfo{journal}{Jpn. J. Appl. Phys} \textbf{\bibinfo{volume}{19}},
  \bibinfo{pages}{L735} (\bibinfo{year}{1980}).

\bibitem[{\citenamefont{Kaufman et~al.}(1999)\citenamefont{Kaufman, Berk, Dwir,
  Rudra, Palevski, and Kapon}}]{kaufman1999conductance}
\bibinfo{author}{\bibfnamefont{D.}~\bibnamefont{Kaufman}},
  \bibinfo{author}{\bibfnamefont{Y.}~\bibnamefont{Berk}},
  \bibinfo{author}{\bibfnamefont{B.}~\bibnamefont{Dwir}},
  \bibinfo{author}{\bibfnamefont{A.}~\bibnamefont{Rudra}},
  \bibinfo{author}{\bibfnamefont{A.}~\bibnamefont{Palevski}}, \bibnamefont{and}
  \bibinfo{author}{\bibfnamefont{E.}~\bibnamefont{Kapon}},
  \bibinfo{journal}{Physical Review B} \textbf{\bibinfo{volume}{59}},
  \bibinfo{pages}{R10433} (\bibinfo{year}{1999}).

\bibitem[{\citenamefont{Oreg et~al.}(2013)\citenamefont{Oreg, Sela, and
  Stern}}]{oreg2013fractional}
\bibinfo{author}{\bibfnamefont{Y.}~\bibnamefont{Oreg}},
  \bibinfo{author}{\bibfnamefont{E.}~\bibnamefont{Sela}}, \bibnamefont{and}
  \bibinfo{author}{\bibfnamefont{A.}~\bibnamefont{Stern}},
  \bibinfo{journal}{arXiv preprint arXiv:1301.7335}  (\bibinfo{year}{2013}).

\bibitem[{\citenamefont{Biscaras et~al.}(2012)\citenamefont{Biscaras, Hurand,
  Feuillet-Palma, Rastogi, Budhani, Reyren, Lesne, LeBoeuf, Proust, Lesueur
  et~al.}}]{biscaras2012irreversibility}
\bibinfo{author}{\bibfnamefont{J.}~\bibnamefont{Biscaras}},
  \bibinfo{author}{\bibfnamefont{S.}~\bibnamefont{Hurand}},
  \bibinfo{author}{\bibfnamefont{C.}~\bibnamefont{Feuillet-Palma}},
  \bibinfo{author}{\bibfnamefont{A.}~\bibnamefont{Rastogi}},
  \bibinfo{author}{\bibfnamefont{R.}~\bibnamefont{Budhani}},
  \bibinfo{author}{\bibfnamefont{N.}~\bibnamefont{Reyren}},
  \bibinfo{author}{\bibfnamefont{E.}~\bibnamefont{Lesne}},
  \bibinfo{author}{\bibfnamefont{D.}~\bibnamefont{LeBoeuf}},
  \bibinfo{author}{\bibfnamefont{C.}~\bibnamefont{Proust}},
  \bibinfo{author}{\bibfnamefont{J.}~\bibnamefont{Lesueur}},
  \bibnamefont{et~al.}, \bibinfo{journal}{arXiv preprint arXiv:1206.1198}
  (\bibinfo{year}{2012}).

\end{thebibliography}

\end{document}